\documentclass[11pt]{article} 
\usepackage{epsfig} 
\newcommand{\sss}{\vspace{.2in}} 
 
\newcommand{\be}{\begin{equation}} 
\newcommand{\ee}{\end{equation}} 
\newcommand{\bea}{\begin{eqnarray}} 
\newcommand{\eea}{\end{eqnarray}} 
\newcommand{\sn}{{\rm sn}} 
\newcommand{\cn}{{\rm cn}} 
\newcommand{\dn}{{\rm dn}} 
\newcommand{\cs}{{\rm cs}} 
\newcommand{\ds}{{\rm ds}}

\newcommand{\dc}{{\rm dc}} 

\newcommand{\ns}{{\rm ns}} 
\newcommand{\nc}{{\rm nc}} 
\newcommand{\Z}{{\rm Z}} 
\newcommand{\sech}{{\rm sech}}

\begin{document} 
\vspace{.2in} 
~\hfill{\footnotesize } 
\vspace{.5in} 
\begin{center} 
{\LARGE {\bf Generalized Landen Transformation 
Formulas for Jacobi Elliptic Functions}} 
\end{center} 
\vspace{.5in} 
\begin{center} 
{\Large{\bf 
   \mbox{Avinash Khare}$^{a,}$\footnote{khare@iopb.res.in} and 
   \mbox{Uday Sukhatme}$^{b,}$\footnote{sukhatme@buffalo.edu} 
 }} 
\end{center} 
\vspace{.6in} 
\noindent 
$^a$Institute of Physics, Sachivalaya Marg, Bhubaneswar 751005, Orissa, India\\ 
$^b$Department of Physics, State University of New York at Buffalo, Buffalo, New York 14260, U.S.A. \\ 
\sss 
\sss 
\vspace{0.7in} 
\begin{center} 
{\Large {\bf Abstract}} 
\end{center} 
Landen transformation formulas, which connect Jacobi elliptic functions with different  
modulus parameters, were first obtained over two hundred years ago by changing integration variables in elliptic integrals. 
We rediscover known results as well as obtain more generalized Landen formulas from a very different perspective,  
by making use of the recently obtained periodic solutions  
of physically interesting nonlinear differential equations and numerous  
remarkable new cyclic identities involving Jacobi elliptic functions.
We find that several of our Landen transformations have a rather different and 
substantially more elegant appearance compared to the forms usually found in the 
literature. Further, by making use of 
the cyclic identities discovered recently, we also
obtain some entirely new sets of Landen transformations.
This paper is an expanded and revised version of our previous paper
math-ph/0204054.

\vspace{.3in}

\newpage 
 
\sss 
  
\section{\bf Introduction}
 
Jacobi elliptic functions $\dn(x,m)$, $\cn(x,m)$ and $\sn(x,m)$, with  
elliptic modulus parameter $m \equiv k^2~(0 \le m \le 1)$ play an important  
role in describing periodic solutions of many linear and nonlinear  
differential equations of interest in diverse branches of engineering,  
physics and mathematics \cite{dj}. 
The Jacobi elliptic functions are often defined 
with the help of the elliptic integral 
\be \label{eq1} 
\int \frac{dz}{\sqrt{(1-z^2)(1-k^2z^2)}}~~.  
\ee 
Over two centuries ago, John Landen \cite{lan}  
studied the consquences of making a change to a different integration variable  
\be\label{eq2}  
t = \frac{(1+k')~z~\sqrt{1-z^2}}{\sqrt{1-k^2z^2}}~~, 
~k' \equiv \sqrt{1-k^2}=\sqrt{1-m}~~.  
\ee
This transformation yields another elliptic integral  
\be\label{eq3} 
\int \frac{dt}{(1+k') \sqrt{(1-t^2)(1-l^2t^2)}}~~,  
~~l \equiv \frac{1-k'}{1+k'}~~.  
\ee  
It readily follows that \cite{han} 
\be\label{eq4} 
\dn\left[(1+k')u, \left(\frac{1-k'}{1+k'}\right)^2\right] 
=\frac{1-(1-k')~\sn^2(u,m)}{\dn(u,m)}~,  
\ee
\be\label{eq5} 
\cn\left[(1+k')u, \left(\frac{1-k'}{1+k'}\right)^2\right] 
=\frac{1-(1+k')~\sn^2(u,m)}{\dn(u,m)}~,\\ 
\ee 
\be\label{eq6}  
\sn\left[(1+k')u, \left(\frac{1-k'}{1+k'}\right)^2\right]= 
\frac{(1+k')~\sn(u,m)~\cn(u,m)}{\dn(u,m)}~.\\ 
\ee   
These celebrated relations are known as the quadratic 
Landen transformation formulas, or more simply, Landen transformations. 
They have  
the special property of providing a non-trivial connection between Jacobi  
elliptic functions involving two different unequal elliptic modulus parameters $m$ and $\tilde {m}$, where 
the transformed modulus parameter is $\tilde {m} \equiv (1-\sqrt{1-m})^2/(1+\sqrt{1-m})^2$. It may be noted that in 
Eqs. (\ref{eq4}) to (\ref{eq6}), 
for $0 < m < 1, ~\tilde{m}$ is always less than $m$ and also lies in the range 
$0 \le \tilde{m} \le 1$. 

There are also similar 
formulas, known as the quadratic Gauss transformations \cite{abr,gra}, 
where the transformed parameter $\tilde{m}$ is always greater than $m$.  
These are obtained by a different change of integration variables
\be\label{eq7}  
t = \frac{(1+k)z}{1+kz^2}~~. 
\ee
This transformation yields the elliptic integral  
\be\label{eq8} 
\int \frac{dt}{(1+k) \sqrt{(1-t^2)(1-l^2t^2)}}~~,  
~~l \equiv \frac{2\sqrt{k}}{1+k}~~,  
\ee  
from which it follows that \cite{han} 
\be\label{eq9} 
\dn\left[(1+k)u, \frac{4k}{(1+k)^2} \right] 
=\frac{1-k\,\sn^2(u,m)}{1+k\,\sn^2(u,m)}~,\\ 
\ee  
\be\label{eq10} 
\cn\left[(1+k)u, \frac{4k}{(1+k)^2} \right] 
=\frac{\cn(u,m)\,\dn(u,m)}{1+k\,\sn^2(u,m)}~,\\ 
\ee 
\be\label{eq11}  
\sn\left[(1+k)u, \frac{4k}{(1+k)^2} \right]= 
\frac{(1+k)\,\sn(u,m)}{1+k\,\sn^2(u,m)}~.\\ 
\ee
In these
Gauss transformation formulas, note that for any choice of $m$ in the range $0 < m <1$, the transformed modulus parameter 
$\tilde{m}=4\sqrt{m}/(1+\sqrt{m})^2$ is always greater
than $m$ but lies in the range $0 < \tilde{m} <1$. 

The above described Landen and Gauss transformation formulas are of order two. 
Subsequently, generalizations of these formulas to arbitrary order $p$ have been studied \cite{lau}. 
The purpose of  this paper is to obtain these generalized transformations
by an entirely different method. In particular, we 
have recently shown \cite{ks1,cks} that a kind of superposition principle works for several 
nonlinear problems like $\lambda \phi^4$ theory and for numerous nonlinear equations of physical interest [Korteweg deVries equation, modified Korteweg deVries equation, nonlinear Schro\"dinger equation, sine-Gordon equation]. Using the idea of superposition, we have obtained seemingly new periodic  
solutions of these nonlinear problems in terms of Jacobi elliptic functions. The purpose of this paper is to argue on mathematical as well as physical grounds that
these solutions cannot really be new, but must be re-expressions of known periodic solutions \cite{rein}. In the process of proving this, we discover the generalized Landen and Gauss transformations. 

As an illustration, let us focus on Eq. (\ref{eq4}) first. Using the identity $\dn^2(u,m)=1-m~\sn^2(u,m)$, and changing variables to $x = (1+k')u$, one can re-write the order two ``$\dn$" Landen formula (\ref{eq4}) in the alternative form 
\be\label{eq12} 
\dn\left[x, \left(\frac{1-k'}{1+k'}\right)^2\right] 
=\alpha ~\left\{\dn \left[\alpha x,m \right]+\dn \left[\alpha x+K(m),m\right]\right\}~,~\alpha=\frac{1}{(1+k')}. 
\ee 
Here, the right hand side contains the sum of two terms with arguments  
separated by $K(m) \equiv \int_0^{\pi/2} d\theta [1-m\sin^2 \theta]^{-1/2}$, 
the complete elliptic integral of the first 
kind \cite{abr,gra}. Our generalized Landen formulas will have not two but $p$ terms on the 
right hand side. We will show that  
the generalization of Eq. (\ref{eq4}) [or equivalently Eq. (\ref{eq12})], valid for any integer $p$   
is given by 
\be\label{eq13} 
\dn (x,{\tilde m}) 
=\alpha \sum_{j=1}^{p} \dn [\alpha x+2(j-1)K(m)/p,m]~,
\ee 
where 
\be\label{eq14} 
\alpha \equiv \left\{ \sum_{j=1}^{p} \dn[2(j-1)K(m)/p,m] \right\}^{-1} ~, 
\ee
and 
\be\label{eq15} 
{\tilde m} =(m-2){\alpha^2}+2{\alpha^3} \sum_{j=1}^{p} \dn^3[2(j-1)K(m)/p,m]~.
\ee

However, if one starts from the ``$\cn$'' or ``$\sn$'' Landen formulas of order 2 as 
given by Eqs. (\ref{eq5}) and (\ref{eq6}), the generalization to arbitrary order $p$ is different 
depending on whether $p$ is even or odd. The main results are:
\bea\label{eq16}
\cn(x,{\tilde m}) \!\!\!\!\!\!&&\propto ~ \sum_{j=1}^p \cn({\tilde x}_j,m)~,~~~~~~~~~~~~~p~ {\rm odd},\nonumber\\
&&\propto ~\sum_{j=1}^p (-1)^{j-1} \dn(x_j,m)~,~~p~ {\rm even},
\eea
\bea\label{eq17}
\sn(x,{\tilde m}) \!\!\!\!\!\!&&\propto ~ \sum_{j=1}^p \sn({\tilde x}_j,m)~,~~~~~~~~~~~~~p~ {\rm odd},\nonumber\\
&&\propto ~ \prod_{j=1}^p \sn(x_j,m)~,~~~~~~~~~~~~~p~ {\rm even},
\eea
where $\tilde m$ is as given in Eq. (\ref{eq15}), and we have used the notation
\be\label{eq18}
x_j \equiv \alpha x+2(j-1)K(m)/p~~,~~\tilde{x_j} \equiv \alpha x+4(j-1)K(m)/p~.
\ee
Note that all the above formulas have the same 
non-trivial scaling factor $\alpha$ of the argument $x$, as well as real shifts which 
are fractions of the periods of Jacobi elliptic functions. The richness of the 
generalized results is noteworthy and reflects the many different forms of 
periodic solutions for nonlinear equations which we have recently 
obtained \cite{ks1,cks}. Some formulas involve the sum 
of $p$ terms, the even $p$ ``$\cn$'' formula
involves alternating $+$ and $-$ signs, and the even $p$ ``$\sn$'' formula 
has a product of $p$ terms. In fact, there are also several interesting additional alternative forms for the above results which follow from use of several identities involving Jacobi elliptic functions which we have recently discovered \cite{ks,kls,kls1}. For instance, we will later show that for odd $p$, the Landen formulas for $\dn,
~\cn, ~\sn$ which have been written above as the sum of $p$ terms, can also be written as the product of the same $p$ terms! 

Analogous to the generalized Landen transformations mentioned above, we also obtain generalized Gauss 
transformations in which the parameter
$\tilde{m}$ is greater than $m$. In particular, it is shown that Gauss transformations can be systematically obtained from the Landen transformations of arbitrary order $p$  
by considering shifts in the arguments by pure imaginary amounts and making use of the 
fact that the Jacobi elliptic functions are doubly periodic.

By starting from the generalized Landen transformations discussed above, we show how to obtain 
Landen transformation formulas for other combinations of Jacobi elliptic functions like 
$\sn(x,\tilde{m}) \cn(x,\tilde{m})$,
$\sn(x,\tilde{m}) \dn(x,\tilde{m})$,
$\cn(x,\tilde{m}) \dn(x,\tilde{m})$,
$\dn^2(x,\tilde{m})$,
$\cn(x,\tilde{m}) \dn(x,\tilde{m}) \sn(x,\tilde{m})$, etc. 
Besides, by integrating the Landen transformation formula for $\dn^2(x,\tilde{m})$ we also
obtain Landen transformations for the Jacobi zeta function $\Z(x,\tilde{m})$ 
as well as for the complete elliptic
integral of the second kind $E(\tilde{m})$.
Finally, by combining the recently discovered cyclic identities 
\cite{ks,kls,kls1} with the 
Landen transformation results, we obtain new types of Landen transformations. For example, a simple
cyclic identity is \cite{kls1}
\be\label{eq19}
\sum_{j=1}^{p} \dn^2 (x_j,m) [\dn(x_{j+r},m)+\dn(x_{j-r},m)] =
2[\ds(a,m)-\ns(a,m)]\sum_{j=1}^{p} \dn(x_j,m)~,
\ee
where 
$a={2rK(m)}/{p}$, $x_j$ is as given by Eq. (\ref{eq18}) and 
$r$ is any integer less than $p$. Combining this
identity with Eq. (\ref{eq13}), we immediately obtain a 
novel Landen
formula relating $\dn(x,\tilde{m})$ with the left hand
side of identity (\ref{eq19}).

The plan of this paper is as follows. In Sec. 2 we obtain solutions of
the sine-Gordon equation in two different ways and by requiring these solutions to be the same, thereby obtain the Landen
transformations for the basic Jacobi elliptic functions $\dn,\cn,\sn$  for both odd and even $p$. 
In Sec. 3 we show that these Landen transformations can be expressed in an
alternative form. Further, we also show that for a given $p$, the relationship
between $\tilde{m}$ and $m$ is the same, regardless of the type of Landen
formula that one is considering.
Using these results, in Sec. 4 we obtain Landen transformations for other
combinations of Jacobi elliptic functions like $\sn \, \cn,\, \sn \,\dn$, etc. both when $p$ is an odd or an even 
integer. In Sec. 5 we obtain Gauss transformations of arbitrary order in
which the parameter $\tilde{m}$ is always greater than $m$. This is done by
starting from the Landen formulas and considering shifts by  
pure imaginary  amounts instead of real shifts. 
In Sec. 6 we combine the recently obtained cyclic identities with
the Landen formulas, thereby obtaining some unusual new Landen transformations. 
Finally, some concluding remarks are given in Sec. 7.

\noindent 

\section{\bf Generalized Landen Formulas} 

Given the diversity of the 
generalized Landen formulas for Jacobi elliptic functions, it is necessary to establish them one 
at a time. 

\subsection{\bf ``dn'' Landen Formulas}

To get an idea of our general approach, let us first focus on the proof of 
Eq. (\ref{eq13}) in detail.  
Consider the periodic solutions of  
the static sine-Gordon field theory in one space and one time dimension, 
that is, consider the periodic solutions of the second order differential equation 
\be\label{eq20} 
\phi_{xx} = \sin \phi~. 
\ee 
Note that the time dependent 
solutions are easily obtained from the static ones by Lorentz boosting.  
One of the simplest periodic solutions of Eq. (\ref{eq20}) is  
given by 
\be\label{eq21} 
\sin [\phi (x)/2] = \dn (x+x_0,{\tilde m})~, 
\ee 
where $x_0$ is an arbitrary constant 
and this is immediately verifiable by direct substitution. It was shown in refs. \cite{ks1,cks}  
that a kind of linear superposition principle works even 
for such nonlinear equations as a consequence of several highly nontrivial, new  
identities satisfied by Jacobi elliptic functions \cite{ks,kls,kls1}. 
In particular, one can 
show \cite{cks} that for any   
integer $p$, one has static periodic solutions of Eq. (\ref{eq20})  
given by 
\be\label{eq22} 
\sin [\phi (x)/2] =\alpha \sum_{j=1}^{p} 
\dn [\alpha x+2(j-1)K(m)/p,m]~, 
\ee 
where $\alpha$ 
is as given by Eq. (\ref{eq14}). 
 
The question one would like to address here is whether solution (\ref{eq22}) is 
completely new, or if it can  
be re-expressed in terms of simpler solutions like (\ref{eq21}), but where $m$ 
and ${\tilde m}$ need not be the same. To that end, consider Eq. (\ref{eq20}) more generally. We note that on 
integrating once, we obtain  
\be\label{eq23} 
\phi_x^2 = C -2\cos \phi~, 
\ee 
where $C$ is a constant of integration. Integrating again, one gets
\be\label{eq24} 
\int \frac{d\phi}{\sqrt{C-2 \cos \phi}} = x+x_0~,  
\ee 
where $x_0$ is a second constant of integration, which we put equal to zero without 
loss of generality since it corresponds to a choice of the origin of coordinates. On substituting  
$\sin (\phi /2) = \psi$,
equation (\ref{eq24}) takes the form 
\be\label{eq25} 
\int \frac{d\psi}{\sqrt{1-\psi^2}\sqrt{\frac{C-2}{4} +\psi^2}} =x~. 
\ee 
Now the important point to note is that if we perform the integral for  
different values of $C$ then we will get 
all the solutions of Eq. (\ref{eq20}). 
Further, if  
two solutions have the same value of $C$, 
then they must necessarily be the same. As far as 
the integral (\ref{eq25}) is concerned, it is easily checked that the three 
simplest solutions covering the entire allowed range of 
$C$ are  
\be\label{eq26} 
\psi = \sech ~x~, ~~C = 2~, 
\ee 
\be\label{eq27} 
\psi = \dn(x,{\tilde m})~, ~~C = 4{\tilde m} -2~, 
\ee 
\be\label{eq28} 
\psi = \cn({x}/{\sqrt{{\tilde m}}}\,,{\tilde m})~,~~C = \frac{4}{{\tilde m}}-2~,  
\ee 
where $0 \le {\tilde m} \le 1$. Note that the constant $C$ has been computed 
here by using Eq. (\ref{eq23}), which in terms of $\psi(x)$ takes the form 
\be\label{eq29} 
C = 2 -4\psi^2 +\frac{4\psi_x^2}{1-\psi^2}~. 
\ee 
Thus, whereas for the solution (\ref{eq27}), $C$ lies in the 
range $-2 \le C \le 2$, for 
the solution (\ref{eq28}), $C$ lies between 2 and $\infty$. Note that for  
$C < -2$, there is no real solution to Eq. (\ref{eq25}). 
 
Now the strategy is clear. We will take the solution (\ref{eq22}) and 
compute $C$ for it and thereby try 
to relate it to one of the basic solutions 
as given by Eqs. (\ref{eq26}) to (\ref{eq28}). One simple way 
of obtaining the constant $C$   
from Eq. (\ref{eq29}) is to evaluate it at a convenient value of $x$, 
say  $x=0$. In this way, we find  
that for the solution  
(\ref{eq22}), $C$ is given by 
\be\label{eq30} 
C = -2 + 4 \alpha^2(m-2)+8\alpha^3 \sum_{j=1}^{p} 
\dn^3 (2(j-1)K(m)/p,m)~~.  
\ee 
Now, as $m \rightarrow 0$, 
$\alpha = 1/p,~ \dn (x,m=0) =1$ and hence $C=-2$. 
On the other hand, as 
$m \rightarrow 1, ~K(m=1) = \infty,~ \dn (x,m=1) = \sech ~x$ and hence  
$\alpha = 1$ so that $C =2$. Thus for solution (\ref{eq22}), 
as $m$ varies 
in the range  
$0 \le m \le 1$, the value of $C$ varies in the range $~-2 \le C \le 2$. 
Hence it is clear that the solutions 
(\ref{eq22}) and (\ref{eq27}) must be same. On equating the two $C$ values as  
given by Eqs. (\ref{eq27}) and (\ref{eq30}), we find that the two solutions 
are identical provided $m$ and ${\tilde m}$ are related by Eq. (\ref{eq15}) 
and hence the appropriate Landen transformation valid for any integer  
$p$ is given by Eq. (\ref{eq13}).  
Note that when $p=2$, one recovers the Landen formula (\ref{eq12}), 
since $\dn(K(m),m)= k'$, and ${\tilde m}$ simplifies to $(1-k')^2/(1+k')^2$.    
 
\noindent 

\subsection{\bf ``cn'' Landen Formulas}  

Unlike the $\dn$ case, it turns out that in this case the Landen transformation formulas
for odd and even $p$ have very different forms. We first derive the form for 
the odd $p$ case and then consider the even $p$ case.

As shown in ref. \cite{cks}, another periodic solution of the static   
sine-Gordon equation (\ref{eq20}) is  
\be\label{eq31} 
\sin (\phi (x)/2) =  \alpha_1 \sum_{j=1}^{p} 
\cn \left [\frac{\alpha_1 x}{\sqrt{m}}+{4(j-1)K(m)}/{p},m \right ]~,  
~~~~p~{\rm odd}~,  
\ee 
with $\alpha_1$ being given by   
\be\label{eq32} 
\alpha_1 \equiv \left\{ \sum_{j=1}^{p} \cn[{4(j-1)K(m)}/{p},m] \right\}^{-1} ~.
\ee 
Using Eq. (\ref{eq29}), we can now compute the corresponding value of the constant $C$. 
We obtain  
\be\label{eq33} 
C = -2 + \frac{4 \alpha_1^2 (1-2m)}{m}+8\alpha_1^3 \sum_{j=1}^{p} \cn^3 [4(j-1)K(m)/p,m]~.  
\ee 
It is easily checked that since $0 \le m \le 1$, $C$ varies from 2 to  
$\infty$, and hence the solutions (\ref{eq28}) and (\ref{eq31}) 
must be identical.  
On equating the two values of $C$ as given by Eqs. (\ref{eq28}) 
and (\ref{eq33}),  
we then find that for odd $p$, the Landen transformation is  
\be\label{eq34} 
\cn (x,{\tilde m_1}) 
=\alpha_1 \sum_{j=1}^{p} \cn \left [\frac{\alpha_1 \sqrt{\tilde{m_1}} x}{\sqrt{m}}+4(j-1)K(m)/p,m \right]~,
\ee 
where 
$\tilde{m_1}$ is given by
\be\label{eq35} 
~~{\tilde m_1} 
= \frac{m}{\alpha_1^2}\left \{(1-2m)+2m\alpha_1 \sum_{j=1}^{p} \cn^3[4(j-1)K(m)/p,m] \right \}^{-1}~.
\ee  

What happens if $p$ is an even integer? As shown in \cite{cks}, 
in that case   
another periodic solution of the static   
sine-Gordon Eq. (\ref{eq20}) is  
\be\label{eq36} 
\sin (\phi (x)/2) =  \alpha_2 \sum_{j=1}^{p} (-1)^{j-1} 
\dn [\alpha_2 x+2(j-1)K(m)/p,m]~,  
~~~~p~{\rm even}~, 
\ee 
 where $\alpha_2$ is given by  
\be\label{eq37}  
\alpha_2 \equiv \left\{ \sum_{j=1}^{p} (-1)^{j-1} \dn[2(j-1)K(m)/p,m] \right\}^{-1} ~.  
\ee 
Using Eq. (\ref{eq29}), the value  
of $C$ for this solution is easily computed and we find that $2 \le C \le   
\infty$. On comparing with solution (\ref{eq28}) we find that in this 
case the  
Landen formula is 
\be\label{eq38} 
\cn (x,{\tilde m_2}) 
=\alpha_2 \sum_{j=1}^{p} (-1)^{j-1} 
\dn [\alpha_2 \sqrt{\tilde{m_2}}~x+2(j-1)K(m)/p,m]~,  
\ee 
where ${\tilde m_2}$ is given by  
\be\label{eq39} 
{\tilde m_2} = \frac{1}{\alpha_2^2}\left \{{(m-2) +2\alpha_2 \sum_{j=1}^{p} (-1)^{j-1} \dn^3 [2(j-1)K(m)/p,m]} \right \}^{-1}~.  
\ee  
As expected, 
in the special case of $p=2$, we immediately recover the Landen 
formula (\ref{eq5}). 

\noindent 

\subsection{\bf ``sn'' Landen Formulas} 

As in the $\cn$ case, here too the Landen formulas for even and odd
values of $p$ have very different forms and so we consider them separately.

We start from  
the sine-Gordon field equation   
\be\label{eq40}  
\phi_{xx}- \phi_{tt}  = \sin \phi~,  
\ee 
and look for time-dependent, traveling wave solutions with velocity $v > 1$ (which are called optical soliton solutions in the context of condensed matter physics).  
In terms of the variable   
\be\label{eq41} 
\eta \equiv \frac{x-vt}{\sqrt{v^2 -1}}~,   
\ee 
Eq. (\ref{eq40}) takes the simpler form  
\be\label{eq42}  
\phi_{\eta \eta} = - \sin \phi~.  
\ee 
Note that the only change from Eq. (\ref{eq20}) is an additional negative sign on the right hand side. On integrating this equation once, we obtain  
\be\label{eq43} 
\phi_{\eta}^2 = C +2\cos \phi~.  
\ee 
On integrating further, we get   
\be\label{eq44} 
\int \frac{d\phi}{\sqrt{C+2 \cos \phi}} = \eta + \eta_0~,   
\ee 
where $\eta_0$ is a constant of integration which we put 
equal to zero without  
loss of generality. Substituting $\sin (\phi /2) = \psi$, yields  
\be\label{eq45} 
\int \frac{d\psi}{\sqrt{1-\psi^2}\sqrt{\frac{C+2}{4} -\psi^2}} = \eta~.  
\ee 
If we now perform the integral for   
different values of $C$, then we get all the solutions.   
It is easily checked that the three  
simplest solutions of Eq. (\ref{eq45}) covering the entire allowed range of  
$C$ are  
\be\label{eq46} 
\psi = \tanh \eta~, ~~C = 2~,  
\ee 
\be\label{eq47} 
\psi = \sqrt{{\tilde m}}   
\, \sn\,(\eta,{\tilde m})~, ~~C = 4{\tilde m} -2~,  
\ee 
\be\label{eq48} 
\psi = \sn\,(\frac{\eta}{\sqrt{{\tilde m}}},{\tilde m})~,~~C = \frac{4}{{\tilde m}}-2~,   
\ee 
where $0 \le {\tilde m} \le 1$.   
Note that the constant $C$ has been computed  
here using Eq. (\ref{eq43}), which in terms of $\psi$ takes the form  
\be\label{eq49} 
C = -2 +4\psi^2 +\frac{4\psi_{\eta}^2}{1-\psi^2}~.  
\ee 
Thus, for solution (\ref{eq47}), $C$ is in the range $-2 \le C \le 2$, whereas for solution (\ref{eq48}), $C$ lies between 2 and $\infty$. Note that for   
$C < -2$, there is no real solution to Eq. (\ref{eq45}).  
  
Using appropriate linear superposition, it was shown in ref.   
\cite{cks} that for odd $p$ one of the solutions of Eq. (\ref{eq42})  
is given by 
\be\label{eq50} 
\sin (\phi (\eta)/2) =  \sqrt{m} \alpha \sum_{j=1}^{p} 
\sn [\alpha \eta+{4(j-1)K(m)}/{p},m]~,  
~~~~p~{\rm odd}~, 
\ee 
with $\alpha$ being given by Eq. (\ref{eq14}).  
Using Eq. (\ref{eq49}), we can now compute the corresponding value of $C$.   
We find 
\be\label{eq51} 
C = -2 + \frac{4 m\alpha^2}{\alpha_1^2}~,  
\ee 
where $\alpha,\alpha_1$ are given by Eqs. (\ref{eq14}) and (\ref{eq32})   
respectively.  
It is easily checked that since 
$0 \le m \le 1$, $C$ has values between -2 and  
2. Hence the solutions (\ref{eq47}) and (\ref{eq50}) must be identical.  
On equating the two values of $C$ 
as given by Eqs. (\ref{eq47}) and (\ref{eq51}),  
we then find that the ``sn" Landen transformation formula for odd $p$   
is given by 
\be\label{eq52} 
\sn (x,{\tilde m_3}) 
= \alpha_1 \sum_{j=1}^{p} \sn [\alpha x+4(j-1)K(m)/p,m]~,  
\ee 
with ${\tilde m_3}$ and $m$ being related by
\be\label{eq53}
\tilde{m_3} = m\frac{\alpha^2}{\alpha_1^2}~.
\ee 
 
Finally, we turn to the ``sn'' 
Landen transformation formula for the case when $p$ is  
an even integer. 
One can show \cite{cks} that in this case, a solution to Eq. (\ref{eq42})   
is given by 
\be\label{eq54} 
\sin (\phi (\eta)/2) =  m^{p/2} \alpha A_0 \prod_{j=1}^{p}   
\sn [\alpha \eta+{2(j-1)K(m)}/{p},m]~,  
~~~~p~{\rm even}~, 
\ee 
with $\alpha$ being given by Eq. (\ref{eq14}) and $A_0$ defined by   
\be\label{eq55} 
A_0 = \prod_{j=1}^{p-1} \sn (2jK(m)/p,m)~.  
\ee 
Using Eq. (\ref{eq49}), we can now compute the corresponding 
value of $C$. We obtain  
\be\label{eq56} 
C = -2 + 4 m^{p} \alpha^4 A_0^4~.  
\ee 
It is easily checked that since $0 \le m \le 1$, the value 
of $C$ varies between -2 and  
2 and hence the solutions (\ref{eq47}) and (\ref{eq54}) must be identical.  
On equating the two values of $C$ as given by Eqs. (\ref{eq47}) 
and (\ref{eq56}),  
we find that for even $p$, the Landen transformation formula   
is 
\be\label{eq57} 
A_0 \,\alpha \,\sn (x,{\tilde m_4})  
= \prod_{j=1}^{p} \sn [\alpha x+2(j-1)K(m)/p,m]~,  
\ee 
with ${\tilde m_4}$ given by  
\be\label{eq58} 
{\tilde m_4} = m^{p} \alpha^4 A_0^4~.  
\ee  
Not surprisingly, for $p=2$ we recover the Landen transformation formula
(\ref{eq6}).
It is amusing to notice that as $m \rightarrow 0$,   
\be\label{eq59} 
A_0 (p,m=0) = \prod_{j=1}^{p-1} \sin (j\pi /p)   
= \frac{p}{2^{p-1}}~.  
\ee 
 
At this point, we have 
generalized all three of the celebrated two hundred year old   
$p=2$ Landen formulas [Eqs. (\ref{eq4}), (\ref{eq5}), (\ref{eq6})] to arbitrary 
values of $p$, the generalization being different depending on whether $p$ 
is an even or odd integer. In the next section, we re-cast the formulas in even simpler form. 
Although several of these Landen formulas 
are already known \cite{lau}, to our knowledge, they have never been derived via the novel approach 
of this article which makes use of solutions of nonlinear field equations. 

\section{\bf Alternative Forms for Landen Transformations}

\subsection{Relation between the transformed modulus parameters $\tilde{m_i}$ and $m$}

So far, we have obtained several seemingly  
different relationships between the transformed modulus parameters $\tilde{m}, \tilde{m_1}, \tilde{m_2}, \tilde{m_3}, \tilde{m_4}$ and
$m$. Let us recall that while relation (\ref{eq13}) for $\tilde{m}$ is valid for all $p$, relations (\ref{eq34}) and (\ref{eq52}) for $\tilde{m_1}$ and $\tilde{m_3}$ are only valid for odd p and relations (\ref{eq38}) and (\ref{eq57}) for $\tilde{m_2}$ and $\tilde{m_4}$ are only valid  for even $p$.  One would like to know if for any given $p$, the different Landen 
transformations give the same relationship between $\tilde{m_i}$ and $m$ or not.
For $p=2$, it is known that $\tilde{m}=\tilde{m_2}=\tilde{m_4}~$ [see Eqs. (\ref{eq4}) to
(\ref{eq6})].
Similarly, with a little algebraic manipulation, one can show
that for $p=3$ the corresponding relations are
\be\label{eq60} 
{\tilde m} = \tilde{m_1} = \tilde{m_3} = m\frac{(1-q)^2}{(1+q)^2(1+2q)^2}~,  
\ee 
where $q \equiv \dn(2K(m)/3,m)$. Note that while deriving this result,   
use has been made of the fact that $\cn(4K(m)/3,m) = -\frac{q}{1+q}$  
and that $q$ satisfies the identity  
$q^4+2q^3-2(1-m)q-(1-m)=0$. 
Similarly, using the relations $\dn (K(m)/2,m) = \dn(3K(m)/2,m) = (1-m)^{1/4}\equiv t$  
and $\dn (K(m),m) = t^2$, it is easily  
proved that for $p=4$, the relations are
${\tilde m} = \tilde{m_2} = \tilde{m_4} = {(1-t)^4}/{(1+t)^4}~$.  

As $p$ increases, the algebra becomes messier and it is not easy to write the
corresponding relations between $\tilde{m_i}$ and $m$ in a neat closed form. However, we can still establish the equivalence of all $\tilde{m_i}$ for any given $p$. For this purpose 
we equate the periods of the left and right hand sides of the various
Landen transformation relations [(\ref{eq13}), (\ref{eq34}), (\ref{eq38}), (\ref{eq52}), (\ref{eq57})]. We get:
\be\label{eq61}
K(\tilde{m}) = \frac{K(m)}{p\alpha}~,
\ee
\be\label{eq62}
K(\tilde{m_1}) = \frac{K(m) \sqrt{m}}{p\alpha_1 \sqrt{\tilde{m_1}}}~,
\ee
\be\label{eq62a}
K(\tilde{m_2}) = \frac{K(m)}{p\alpha_2 \sqrt{\tilde{m_2}}}~,
\ee
\be\label{eq62b}
K(\tilde{m_3}) = \frac{K(m)}{p\alpha}~,
\ee
\be\label{eq62c}
K(\tilde{m_4}) = \frac{K(m)}{p\alpha}~.
\ee
>From Eqs. (\ref{eq61}), (\ref{eq62b}) and (\ref{eq62c}) above, one immediately sees that $\tilde{m} = \tilde{m_3}$ for odd $p$ and $\tilde{m} = \tilde{m_4}$ for even $p$. In order to establish that $\tilde{m_1} = \tilde{m_3}$ for odd $p$, we suitably rescale, square and add the Landen transformations (\ref{eq34}) and (\ref{eq52}). This yields
\be \label{eq62d}
\sn^2(x,\tilde{m}) + \cn^2(\frac{\alpha \sqrt{m}x}{\alpha_1 \sqrt{\tilde{m_1}}},\tilde{m_1}) = C~,
\ee
where the constant $C$ on the right hand side comes from making use of the cyclic identities \cite{ks,kls,kls1} given in eqs. (\ref{eq97}) and (\ref{eq98}).
Since Eq. (\ref{eq62d}) is valid for all $x$, and in particular for $x=0$, it follows that $C=1$, and hence one must have
\be
\sn^2(x,\tilde{m}) = \sn^2(\frac{\alpha \sqrt{m}x}{\alpha_1 \sqrt{\tilde{m_1}}},\tilde{m_1}).
\ee
This implies that $\frac{\alpha \sqrt{m}}{\alpha_1 \sqrt{\tilde{m_1}}}=1$ and hence $\tilde{m_1}=\tilde{m}$ for odd $p$. Similar reasoning yields 
$\frac{\alpha}{\alpha_2 \sqrt{\tilde{m_2}}}=1$ and $\tilde{m_2}=\tilde{m}$ for even $p$. 

Consequently, for odd $p$, the three Landen transformations are
\be\label{eq64}
\dn(x,\tilde{m}) = \alpha \sum_{j=1}^{p} 
\dn (x_j,m)~,
\ee
\be\label{eq65}
\cn(x,\tilde{m}) = \alpha_1 \sum_{j=1}^{p} 
\cn (\tilde{x_j},m)~,
\ee
\be\label{eq66}
\sn(x,\tilde{m}) = \alpha_1 \sum_{j=1}^{p} 
\sn (\tilde{x_j},m)~,
\ee
where $\alpha,\alpha_1$ are given by Eqs. (\ref{eq14}) and (\ref{eq32}), 
$x_j,\tilde{x_j}$ are as given by Eq. (\ref{eq18})  while $\tilde{m}$
and $m$ are related by Eq. (\ref{eq15}).

Likewise, for even $p$, the three Landen formulas are
\be\label{eq68}
\dn(x,\tilde{m}) = \alpha \sum_{j=1}^{p} 
\dn (x_j,m)~,
\ee
\be\label{eq69}
\cn(x,\tilde{m}) = \alpha_2 \sum_{j=1}^{p} (-1)^{j-1}  
\dn (x_j,m)~,
\ee
\be\label{eq70}
A_0 \,\alpha \,\sn(x,\tilde{m}) =  \prod_{j=1}^{p} 
\sn (x_j,m)~,
\ee
where $\alpha,\alpha_2, A_0$ are given by Eqs. (\ref{eq14}), (\ref{eq37}) 
and (\ref{eq55}) respectively 
while $\tilde{m}$
and $m$ are related by Eq. (\ref{eq15}).

The numerical results for ${\tilde m}$ as a function of $m$ for various 
values of $p$ ranging from $2$ to $7$ are shown in Table 1. 
Note that for any 
fixed value of $p$, as $m$ increases from 0 to 1, 
${\tilde m}$ also increases 
monotonically from 0 to 1 (but is always less than $m$). 
Also, for any given fixed value of $m$, 
${\tilde m}$ decreases monotonically as $p$ increases and 
is always less than
$m$. For this reason, Landen transformations are sometimes referred to  
as ascending Landen 
transformations \cite{abr}.   

\subsection{Alternative Forms of Landen Transformation Formulas}
  
Recently we \cite{ks,kls,kls1} have obtained several new identities for 
Jacobi elliptic functions. Three of these, valid for odd $p$, are
\be\label{eq71}
\prod_{j=1}^{p} \dn(x_j,m) = \prod_{n=1}^{(p-1)/2} \cs^2(\frac{2Kn}{p},m) 
\sum_{j=1}^{p} \dn(x_j,m)~,
\ee
\be\label{eq72}
 \prod_{j=1}^{p} \sn(\tilde{x}_j,m) 
= (-1/m)^{(p-1)/2} \prod_{n=1}^{(p-1)/2} \ns^2(\frac{4Kn}{p},m) 
\sum_{j=1}^{p} \sn(\tilde{x}_j,m)~,
\ee
\be\label{eq73}
 \prod_{j=1}^{p} \cn(\tilde{x}_j,m) 
= (1/m)^{(p-1)/2} \prod_{n=1}^{(p-1)/2} \ds^2(\frac{4Kn}{p},m) 
\sum_{j=1}^{p} \cn(\tilde{x}_j,m)~.
\ee
Here $\ds(x,m)$, etc. are defined by $\ds(x,m)=\frac{\dn(x,m)}{\sn(x,m)}$. 
Hence, for odd $p$, the three Landen transformation formulas can
also be written as 
products (rather than sums) of $p$ terms. 
In particular, for odd $p$ the three Landen transformation
formulas (\ref{eq64}) to (\ref{eq66}) can also be written in the form
\be\label{eq75}
\dn(x,\tilde{m})
= \frac{{\prod_{j=1}^{p}} {\dn (x_j,m)}}{\prod_{n=1}^{p-1} \dn(2nK(m)/p,m)}~,
\ee
\be\label{eq76}
 \cn(x,\tilde{m})
= \frac{\prod_{j=1}^{p} \cn (\tilde{x}_j,m)}{\prod_{n=1}^{p-1} \cn(4nK(m)/p,m)}~,
\ee
\be\label{eq77}
\sn(x,\tilde{m})
= \frac{\prod_{j=1}^{p} \sn (\tilde{x}_j,m)}
{\alpha(m) \prod_{n=1}^{p-1} \sn(4nK(m)/p,m)}~.
\ee

Similarly, for even $p$, on making use of the identity \cite{ks,kls}
\be\label{eq79}
(m)^{p/2} \prod_{j=1}^{p} \sn (x_j,m)
= \big [\prod_{n=1}^{\frac{p}{2}-1} \ns^2(2nK(m)/p,m) \big ] 
\sum_{j=1}^{p} (-1)^{j-1} \Z \big (x_j,m \big ),
\ee
where $\Z(u,m)$ is the Jacobi zeta function, 
we can rewrite the Landen formula (\ref{eq70}) for 
$\sn$ in the form
\be\label{eq80}
\sn(x,\tilde{m})=
\alpha_2 \sum_{j=1}^{p} (-1)^{j-1} \Z (x_j,m)~.
\ee
We have not seen this particular form for the $\sn$ Landen
transformation in the mathematics literature. Here the coefficient multiplying
the right hand side of Eq. (\ref{eq80}) has been fixed by demanding 
consistency. From now onward, we shall mostly be using this form of the
Landen formula rather than the one given by Eq. (\ref{eq70}). 

On comparing Eqs. (\ref{eq70}), (\ref{eq79}) and (\ref{eq80}) we obtain an
interesting identity
\be\label{eq81}
m^{p/2} \alpha \alpha_2 \prod_{j=1}^{p-1} \sn (2jK(m)/p,m)
= \prod_{n=1}^{\frac{p}{2}-1} \ns^2(2nK(m)/p,m)~. 
\ee

In concluding this section, we note that many of the above derived Landen transformation formulas lead to interesting known  trigonometric relations by taking the limiting case $m = \tilde{m}=0$. For instance, Eqs. (\ref{eq70}), (\ref{eq76}) and (\ref{eq77}) lead to \cite{gra}
\be
\sin px = 2^{p-1} \prod_{j=1}^{p} \sin[x+(j-1)\pi/p]~,~~ p~ {\rm even},
\ee
\be
\cos px = 2^{p-1} \prod_{j=1}^{p} \cos[x+2(j-1)\pi/p]~, ~~p~ {\rm odd},
\ee
\be
\sin px = (-4)^{\frac{p-1}{2}} \prod_{j=1}^{p} \sin[x+2(j-1)\pi/p]~,~~ p~ {\rm odd}.
\ee

\section{\bf Landen Transformations for Products of Jacobi Elliptic Functions}

We shall now show that starting from the Landen formulas for the three
basic Jacobi elliptic functions $\sn,~\cn,~\dn$, we can obtain Landen formulas for 
their products and various other
combinations.

\subsection{Any integer $p$}

We start from the basic Landen formula, Eq. (\ref{eq64}) 
 valid for any integer $p$. Differentiating 
it gives the 
Landen formula:
\be\label{eq82}
\sn(x,\tilde{m})\cn(x,\tilde{m})
=\frac{m\alpha^2}{\tilde{m}}\sum_{j=1}^{p} \sn (x_j,m) \cn (x_j,m)~,
\ee
For odd values of $p$, this can also be proved by multiplying 
the two
Landen formulas [Eqs. (\ref{eq65}) and
(\ref{eq66})] and using the cyclic identity 
\be\label{eq85}
\sum_{j=1}^{p} \sn(x_j,m)[\cn(x_{j+r},m)+\cn(x_{j-r},m)] =0~,
\ee
obtained
in refs. \cite{ks,kls}. 
Here $r=1,2,...,(p-1)/2$.
On the other hand, for even $p$, 
multiplying relations (\ref{eq69}) and (\ref{eq80})
and comparing with identity (\ref{eq82}) yields the remarkable identity
\be\label{eq101}
m \sum_{j=1}^{p} \sn (x_j,m)\cn (x_j,m)
= \left[ \sum_{j=1}^{p} (-1)^{j-1} 
\Z (x_j,m) \right] \left[ \sum_{j=1}^{p} (-1)^{j-1} \dn (x_j,m) \right].
\ee

Nontrivial results also follow from squaring any of the Landen formulas. 
For example, squaring the Landen formula (\ref{eq64}) and using
the cyclic identity
\be\label{eq86}
\sum_{j=1}^{p} \dn(x_j,m) \dn(x_{j+r},m)
=p[\dn(a,m)-\cs(a,m)\Z(a,m)]~,~~a \equiv 2rK(m)/p~,
\ee
yields the Landen formula for $\dn^2$, i.e. we get
\be\label{eq87}
\dn^2(x,\tilde{m})
=\alpha^2 \left [\sum_{j=1}^{p} \dn^2 (x_j,m)
+2A_d \right ]~,
\ee
where
\bea\label{eq88}
&&A_d \equiv \sum_{i < j=1}^{p} \dn(x_i,m) \dn(x_j,m) \nonumber \\ 
&&= p \sum_{j=1}^{(p-1)/2}\big [\dn(2jK(m)/p,m)-\cs(2jK(m)/p,m)\Z(2jK(m)/p) \big]~,
~~p~ {\rm odd}~, \nonumber \\
&&= (p/2)\sqrt{1-m} 
+p \sum_{j=1}^{(p-2)/2}\big [\dn(2jK(m)/p,m)-\cs(2jK(m)/p,m)\Z(2jK(m)/p) \big]~,
~p~ {\rm even}.
\eea
It is worth noting that the linearly superposed solutions of KdV equation
\cite{ks1} discovered recently, essentially correspond to this Landen 
transformation. 

Differentiating both sides of 
Eq. (\ref{eq87}) yields the Landen transformation
for the product $\sn~\cn~\dn$ of the three basic Jacobi elliptic functions. We obtain
\be\label{eq89}
\dn(x,\tilde{m})\sn(x,\tilde{m})\cn(x,\tilde{m}) 
=\frac{m\alpha^2}{\tilde{m}}\sum_{j=1}^{p} 
\dn (x_j,m) \sn (x_j,m) \cn (x_j,m).
\ee

On the other hand, if we integrate both sides of (\ref{eq87}) with respect to $x$, then 
we simulteneously obtain Landen transformations for both the Jacobi Zeta function
as well as the complete elliptic integral of the second kind $E(m)$. In particular,
on using the well known formula \cite{abr,gra}
\be\label{x}
\int \dn^2 (x,m)\, dx = \Z (x,m) +\frac{E(m)}{K(m)}x~,
\ee
integration of Eq. (\ref{eq87}) yields 
\be\label{y}
\Z (x,\tilde{m})+\frac{E(\tilde{m})}{K(\tilde{m})}x
=\alpha^2 x \left[2A_d+p\frac{E(m)}{K(m)} \right] 
+\alpha\sum_{j=1}^{p} 
\Z (x_j,m)~.
\ee
Here the constant on the right hand side has been fixed by considering
the equation at $x=0$ and using the fact that $\Z(x=0,m)=0$, that it is an odd
function of its argument $x$ and that it is a periodic function with period 
$2K(m)$. 
It is now obvious that terms proportional to $x$ on the left and right hand sides 
must cancel among themselves, since the remaining terms are oscillatory and do 
not grow. This immediately yields two remarkable Landen transformations:  
\be\label{y1}
\Z (x,\tilde{m})
=\alpha\sum_{j=1}^{p} 
\Z (x_j,m)~,
\ee
\be\label{y2}
E(\tilde{m})
=\alpha \left[E(m)+ \frac{2A_d K(m)}{p}\right],
\ee
where use has been made of relation (\ref{eq61}).
As expected, for $p=2$ the Landen transformation agrees with the well known 
relation given in \cite{abr}. However, it seems that  
the generalized Landen transformation for Jacobi zeta functions is a new result.
 
Landen formulas for 
higher order combinations like say $\dn^n (x,\tilde{m})$ can
be obtained recursively by differentiating the lower order Landen
formulas (\ref{eq82}) and (\ref{eq89}). 
For example, differentiating relation (\ref{eq82}) yields
the Landen formula for $\dn^3(x,\tilde{m})$
\be\label{eq90}
\dn^3(x,\tilde{m})
=\alpha^3 \sum_{j=1}^{p} \dn^3 (x_j,m)
+\alpha[2-\tilde{m}-(2-m)\alpha^2]
\sum_{j=1}^{p} \dn(x_j,m)~.
\ee

\subsection{Odd Integer Case}

We start from the two basic Landen formulas, Eqs. (\ref{eq65}) and (\ref{eq66})
valid for any odd integer $p$. Differentiating and using the relation between
$\tilde{m}$ and $m$, gives the following Landen
formulas:
\be\label{eq83}
\sn(x,\tilde{m})\dn(x,\tilde{m})
=\alpha \alpha_1\sum_{j=1}^{p} \sn (\tilde{x_j},m) \dn (\tilde{x_j},m)~,
\ee
\be\label{eq84}
\cn(x,\tilde{m})\dn(x,\tilde{m})
=\alpha \alpha_1\sum_{j=1}^{p} \cn (\tilde{x_j},m) \dn (\tilde{x_j},m)~.
\ee
It may be noted that these relations can also be proved by multiplying appropriate 
Landen formulas [Eqs. (\ref{eq64}) to
(\ref{eq66})] and using the cyclic identities analogous to (\ref{eq85}) obtained
in refs. \cite{ks,kls}. 

Landen formulas for higher order combinations like  
$\sn^{2n+1}(x,\tilde{m})$ or $\cn^{2n+1}(x,\tilde{m})$ can be obtained 
recursively from here by differentiating the lower order Landen formulas.
For example,  
differentiation of relations (\ref{eq83}) and (\ref{eq84})
yield the following Landen formulas for 
$\sn^3(x,\tilde{m})$ and $\cn^3(x,\tilde{m})$:
\be\label{eq91}
\cn^3(x,\tilde{m})
=\alpha_1^3\bigg [\sum_{j=1}^{p} \cn^3 (\tilde{x_j},m)
+\left (\frac{1-\alpha_1^2}{\alpha_1^2} -\frac{1-\alpha^2}{2m\alpha^2}
\right ) \sum_{j=1}^{p} \cn (\tilde{x_j},m) \bigg ],
\ee
\be\label{eq92}
\sn^3(x,\tilde{m})
=\alpha_1^3\bigg [\sum_{j=1}^{p} \sn^3 (\tilde{x_j},m)
+\left (\frac{1-\alpha^2}{2m\alpha^2} +\frac{1-\alpha_1^2}{2\alpha_1^2}
\right ) \sum_{j=1}^{p} \sn (\tilde{x_j},m) \bigg ].
\ee

Before concluding the discussion about the Landen formulas for odd $p$, we
want to point out the consistency conditions which we obtain by demanding
$\dn^2(x,\tilde{m})+\tilde{m}\sn^2(x,\tilde{m})=1$, and 
$\sn^2(x,\tilde{m})+\cn^2(x,\tilde{m})=1$. In particular, using relations
(\ref{eq64}) to (\ref{eq66}) and demanding these constraints, yields
\be\label{eq93}
\frac{1}{\alpha^2} = p+2(A_d+A_s)~,~~
\frac{m}{\alpha_1^2} = mp+2(A_s+A_c)~,
\ee 
where in view of the cyclic identities derived in \cite{ks,kls,kls1} we have
\be\label{eq94}
A_s \equiv m\sum_{i < j =1}^{p} \sn(x_i,m) \sn(x_j,m) =
p\sum_{j=1}^{(p-1)/2} \frac{\Z(2jK(m)/p,m)}{\sn(2jK(m)/p)}~,
\ee
\be\label{eq95}
A_c \equiv m\sum_{i < j =1}^{p} \cn(x_i,m) \cn(x_j,m) =
p\sum_{j=1}^{(p-1)/2} \bigg [\cn(2jK(m)/p)- 
\frac{\dn(2jK(m)/p,m) \Z(2jK(m)/p,m)}{\sn(2jK(m)/p)} \bigg ],
\ee
and $A_d$ is as given by Eq. (\ref{eq88}).

\subsection{Even Integer Case}

We start from the two Landen formulas for even $p$ given by Eqs. 
(\ref{eq69}) and (\ref{eq80}). Differentiating them and using the
relation for $\tilde{m}$ leads to the following
Landen formulas: 
\be\label{eq97}
\sn(x,\tilde{m})\dn(x,\tilde{m})
=m \alpha \alpha_2\sum_{j=1}^{p} (-1)^{j-1} \sn (x_j,m) \cn (x_j,m),
\ee
\be\label{eq98}
\cn(x,\tilde{m})\dn(x,\tilde{m})
=\alpha \alpha_2\sum_{j=1}^{p} (-1)^{j-1} 
\dn^2(x_j,m).
\ee

It may be noted that the relation (\ref{eq98}) can also be derived by 
multiplying the Landen formulas (\ref{eq68}) and (\ref{eq69}) and using
the cyclic identity 
\be\label{eq99}
[\sum_{j=1}^{p} \dn(x_j,m)]\sum_{j=1}^{p} (-1)^{j-1} \dn(x_j,m)
=\sum_{j=1}^{p} (-1)^{j-1} \dn^2 (x_j,m)~.
\ee
On the other hand, multiplying relations (\ref{eq68}) and (\ref{eq80})
and comparing with identity (\ref{eq97}) yields a remarkable identity
\be\label{eq100}
m \sum_{j=1}^{p} (-1)^{j-1} 
\sn (x_j,m)\cn (x_j,m)= \left [\sum_{j=1}^{p} (-1)^{j-1} 
\Z (x_j,m) \right ]
 \sum_{j=1}^{p} \dn(x_j,m)~.
\ee

Landen formulas for higher powers like say $\sn^{2n+1} (x,\tilde{m})$ can
be obtained recursively from here by differentiating the lower order Landen
formulas. 
For example, differentiation of the relations (\ref{eq97}) and (\ref{eq98})
gives rise to the following Landen formulas for 
$\cn^3(x,\tilde{m})$ and $\sn^3(x,\tilde{m})$:
\be\label{eq106}
\cn^3(x,\tilde{m})
=\alpha_2^3\bigg [\sum_{j=1}^{p} (-1)^{j-1} 
\dn^3 (x_j,m)
+\left(\frac{2\alpha^2-\alpha_2^2}{2\alpha^2 \alpha_2^2} 
+\frac{m-2}{2}
\right) \sum_{j=1}^{p} (-1)^{j-1} 
\dn (x_j,m) \bigg ],
\ee
\be\label{eq107}
\sn^3(x,\tilde{m})
=\alpha_2 \left(\frac{\alpha^2+\alpha_2^2}{2\alpha^2}\right)
\sum_{j=1}^{p} (-1)^{j-1}\Z (x_j,m)
-m \alpha_2^3 
\sum_{j=1}^{p} (-1)^{j-1} 
\sn (x_j,m) \cn (x_j,m) \dn (x_j,m)~.
\ee

Before finishing the discussion about the Landen formulas for even $p$, we
want to point out the consistency conditions which we obtain by demanding
$\dn^2(x,\tilde{m})+\tilde{m}\sn^2(x,\tilde{m})=1$, and 
$\sn^2(x,\tilde{m})+\cn^2(x,\tilde{m})=1$. In particular, using relations
(\ref{eq68}), (\ref{eq69}) and (\ref{eq80}) 
and demanding these constraints, yields
\be\label{eq108}
\frac{1}{\alpha^2} = \dn^2 (x_j,m)
+2A_d + \left [\sum_{j=1}^{p} (-1)^{j-1} \Z (x_j,m) \right ]^2~,
\ee 
\be\label{eq109}
\frac{1}{\alpha_2^2} = 
\left [\sum_{j=1}^{p} (-1)^{j-1} \dn (x_j,m) \right ]^2
+ \left [\sum_{j=1}^{p} (-1)^{j-1} \Z (x_j,m) \right ]^2~,
\ee 
where $A_d$ is as given by Eq. (\ref{eq88}).

\section{Gauss Transformation Formulas}

For $p=2$, one has the Gauss quadratic transformation formulas \cite{abr}
as given by Eqs. (\ref{eq9}) to (\ref{eq11}) in which $\tilde{m}$ is 
greater than $m$. Gauss transformations are sometimes referred to as descending Landen 
transformations \cite{abr}. It is natural to ask whether Gauss transformation formulas can be generalized to 
arbitrary $p$. In this 
context it may be noted   
that the Landen transformation generalizations which we have established so far are in 
terms of shifts involving the period $K(m)$ on the real axis.
We now show that the generalization of
the Gauss transformation formulas to arbitrary order result from 
shifts involving the period $iK'(m)$ on the imaginary axis.

The procedure consists of starting with any Landen transformation formula,
using the standard results \cite{han,abr}
\be\label{e1}
\dn(x,m') = \dc(ix,m)~,~~
\cn(x,m') = \nc(ix,m)~,~~
\sn(x,m') = -i{\rm sc} (ix,m)~,
\ee
and then redefining $ix =u$. Note that $m' = 1-m$ while 
$\dc(x,m) \equiv \frac{\dn(x,m)}{\cn(x,m)}$, etc. In this way, we find that 
for odd $p$, the
three Landen transformation formulas as given by Eqs. (\ref{eq64}) to 
(\ref{eq66}) take the form
\be\label{e2}   
\dc(x,\tilde{m})=\beta \sum_{j=1}^{p} 
\dc [\beta x+2i(j-1)K'(m)/p,m]~,
\ee
\be\label{e3}   
\nc(x,\tilde{m})=\beta_1 \sum_{j=1}^{p} 
\nc [\beta x+4i(j-1)K'(m)/p,m]~,
\ee
\be\label{e4}   
{\rm sc} (x,\tilde{m})
=\beta_1 \sum_{j=1}^{p} 
{\rm sc} [\beta x+4i(j-1)K'(m)/p,m]~,
\ee
where 
\be\label{e5}
\beta \equiv \beta (m) = \alpha (m'=1-m)
=\left \{ {\sum_{j=1}^{p} \dn[2(j-1)K'(m)/p,m']} \right \}^{-1}~,
\ee
\be\label{e6}
\beta_1 \equiv \beta_1 (m) = \alpha_1 (m'=1-m)
=\left \{ {\sum_{j=1}^{p} \cn[4(j-1)K'(m)/p,m']} \right \}^{-1}~.
\ee

It might be noted here that the Landen formula (\ref{e2}) is in fact valid
for both even and odd $p$.
On the other hand, for even $p$, instead of relations (\ref{e3}) and 
(\ref{e4}) we have the Landen formulas  
\be\label{e7}   
\nc(x,\tilde{m})=\beta_2 \sum_{j=1}^{p} (-1)^{j-1} 
\dc[\beta x+2i(j-1)K'(m)/p,m]~,
\ee
\be\label{e8}   
{\rm sc} (\beta x,\tilde{m})\beta B_0= (-i)^{p-1}  
\prod_{j=1}^{p} {\rm sc} [\beta x+2i(j-1)K'(m)/p,m]~,
\ee
where $\beta_2$ and $B_0$ are given by 
\be\label{e9}
\beta_2 \equiv \beta_2 (m) = \alpha_2 (m'=1-m)
= \left \{{\sum_{j=1}^{p} (-1)^{j-1} \dn[2(j-1)K'(m)/p,m']}\right \}^{-1}~,
\ee
\be\label{e10}
B_0 \equiv B_0 (m) = A_0 (m') = \prod_{j=1}^{p-1} \sn [2jK'(m)/p,m']~.  
\ee
For odd $p$, $\tilde{m}$ is now given by
\be\label{e11}
1-\tilde{m} = (1-m)\frac{\beta^2}{\beta_1^2}~,
\ee
while for even $p$ it is given by 
\be\label{e12}
1-\tilde{m} = \frac{\beta^2}{\beta_2^2}~,
\ee
where $\beta,\beta_{1},\beta_2$ are as given by Eqs. (\ref{e5}), (\ref{e6}) and
(\ref{e9}) respectively.

In the special case of $p=2$, it is easily checked that 
$\tilde{m} =4\sqrt{m} /(1+\sqrt{m})^2$  and as expected
the relations
(\ref{e2}),  (\ref{e7}) and (\ref{e8}) reduce to the well known \cite{abr}
Gauss transformation formulas as given by Eqs. (\ref{eq9}) to (\ref{eq11}). Note that there is an obvious, intersting relationship between the transformed modulus parameters $\tilde{m_G}(m)$ and $\tilde{m_L}(m)$ in Gauss and Landen transformations respectively. For any choice of $p$, the Gauss transformation undoes the effects of the Landen transformation and vice versa. This means that $\tilde{m_G}[\tilde{m_L}(m)] = m$ and $\tilde{m_L}[\tilde{m_G}(m)] = m$. For $p=2$, these relations are easily checked from Eqs. (\ref{eq4}) and (\ref{eq9}). 

Before ending this section, it is worth remarking that just as we have 
considered Landen formulas where the shifts are 
in units of $K(m)$ or $iK'(m)$ on the real or
imaginary axis respectively, we can also consider Landen formulas 
corresponding to shifts in units of $K(m)+iK'(m)$ in the complex plane. In 
this context it is worth noting that by 
starting from the Landen formulas for $p=2$ as given by Eqs. 
(\ref{eq4}) to (\ref{eq6}), changing $m$ to $1/m$, $x$ to $kx$, 
and using the formulas
\be\label{e13}
\dn \big (kx,\frac{1}{m} \big ) = \cn(x,m)~,
~~\cn \big (kx,\frac{1}{m} \big ) =\dn(x,m)~,~~
\sn \big (kx,\frac{1}{m} \big ) =k \sn(x,m)~,~~
\ee
one readily obtains the Landen transformations \cite{han}
\be\label{e15a} 
\dn\left[(k+ik')u, \left( \frac{k-ik'}{k+ik'}\right )^2 \right]= 
\frac{1-k(k-ik')~\sn^2(u,m)}{\cn(u,m)}~,\\ 
\ee  
\be\label{e15} 
\cn\left[(k+ik')u, \left( \frac{k-ik'}{k+ik'}\right )^2 \right]= 
\frac{1-k(k+ik')~\sn^2(u,m)}{\cn(u,m)}~,\\ 
\ee 
\be\label{e14}  
\sn\left[(k+ik')u, \left( \frac{k-ik'}{k+ik'}\right )^2 \right]= 
\frac{(k+ik')~\sn(u,m)~\dn(u,m)}{\cn(u,m)}~.\\ 
\ee 

The generalizations of these Landen transformations to arbitrary $p$ is
immediate. In particular, by changing $m (\tilde{m})$ to $1/m (1/\tilde{m})$,
changing $x$ to $kx$  
and using formulas (\ref{e13}) in the Landen formulas
as given by
Eqs. (\ref{eq64}) to (\ref{eq70}), we obtain the corresponding 
Landen formulas 
for shifts by complex amounts in units of $K(m)+iK'(m)$. For example, for
any integer $p$, we get 
\be\label{e16}
\cn(x,\tilde{m})=\delta \sum_{j=1}^{p} 
\cn [\delta_1 x+2(j-1) (K(m)+iK'(m))/p,m]~,
\ee
while for odd $p$ we have
\be\label{e17}
\dn(x,\tilde{m})=\delta_2 \sum_{j=1}^{p} 
\dn [\delta_2 x+4(j-1)(K(m)+iK'(m))/p,m]~,
\ee
\be\label{e18}
\sn(x,\tilde{m})=\delta_1 \sum_{j=1}^{p} 
\sn [\delta_2 x+4(j-1)(K(m)+iK'(m))/p,m]~,
\ee
with $\tilde{m}$ and $m$ being related by 
$\tilde{m}={\delta_1^2}/{\delta^2}.$

On the other hand, the corresponding 
Landen formulas for even $p$ are
\be\label{e19}
\dn(x,\tilde{m})=\delta_2 \sum_{j=1}^{p} (-1)^{j-1} 
\cn \big ({\delta_2 x}/{\sqrt{m}}+2(j-1)(K(m)+iK'(m))/p,m \big )~,
\ee
\be\label{e20}
\delta_2 D_0 \sn(x,\tilde{m})= \sqrt{m} \prod_{j=1}^{p}  
\sn \big ({\delta_2 x}/{\sqrt{m}}+2(j-1)(K(m)+iK'(m))/p,m \big )~,
\ee
where $\tilde{m}$ is given by
$\tilde{m} = {\delta_2^2}/{\delta^2}.$
Here, $\delta,\delta_{1},\delta_2$ are given by
\be\label{e21}
\delta \sum_{j=1}^{p} \cn [2(j-1)(K(m)+iK'(m))/p,m] = 1~,
\ee
\be\label{e22}
\delta_1 \sum_{j=1}^{p} \dn [4(j-1)(K(m)+iK'(m))/p,m] = 1~,
\ee
\be\label{e23}
\delta_2 \sum_{j=1}^{p} (-1)^{j-1} 
\cn [2(j-1)(K(m)+iK'(m))/p,m] = 1~,
\ee
while $D_0$ is given by
\be
D_0 = \prod_{j=1}^{p-1} \sn [2j(K(m)+iK'(m))/p,m]~.
\ee
It is easily checked that for $p=2$ the Landen formulas (\ref{e16}),
(\ref{e19}) and 
(\ref{e20}) reduce to the well known ``complex'' Landen formulas (\ref{e15}), (\ref{e15a})
and (\ref{e14}) respectively.

\section{Landen Transformation Formulas and Cyclic Identities}

Recently, we \cite{ks, kls} have obtained a large number of cyclic identities
where combinations of 
Jacobi elliptic functions at different points are expressed in terms of sums like 
$\sum_{j=1}^{p} \dn(x+2(j-1)K(m)/p,m)$,
$\sum_{j=1}^{p} (-1)^{j-1}\dn(x+2(j-1)K(m)/p,m)$,
etc. Now the remarkable
thing is that it is precisely these sums for which Landen and others
have obtained the famous transformation formulas mentioned above. Thus by 
combining the cyclic identities and the Landen formulas given above, we can
obtain a wide class of {\it generalized} Landen transformations for many
combinations of Jacobi elliptic functions. 

For example, a simple cyclic identity valid for both even and odd
$p$ is
\be\label{f1}
\sum_{j=1}^{p} \dn^2 (x_j,m) [\dn(x_{j+r},m)+\dn(x_{j-r},m)] 
= 2[\ds(a,m)\ns(a,m)-\cs^2(a,m)]\sum_{j=1}^{p} \dn(x_j,m)~,
\ee
where $x_j$  is given by Eq. (\ref{eq18}) 
and $a=2rK(m)/p$. On combining with relation (\ref{eq64}), we obtain the Landen formula for the
left hand side of (\ref{f1}) given by
\be\label{f2}
\sum_{j=1}^{p} \dn^2 (x_j,m) [\dn(x_{j+r},m)+\dn(x_{j-r},m)] 
=(2/\alpha)[\ds(a,m)\ns(a,m)-\cs^2(a,m)] 
\dn (x,\tilde{m})~,
\ee
where $\tilde{m}$ and $m$ are related by Eq. (\ref{eq15}).

Proceeding in this way, using the many identities obtained by us
\cite{ks,kls,kls1}, we obtain a huge class of new Landen transformation formulas. As 
illustrations, we give four examples:
\be\label{f3}
m \sum_{j=1}^{p} \cn(x_j,m) [\sn(x_{j+r},m)-\sn(x_{j-r},m)] = 
(2/\alpha)[\ns(a,m)-\ds(a,m)] 
\dn (x,\tilde{m})~;
\ee
\be\label{f4}
\sum_{j=1}^{p} \dn^2 (x_j,m) [\dn(x_{j+r},m)-\dn(x_{j-r},m)] 
=-(2m/A)\cs(a,m) \cn (x,\tilde{m}) \sn (x,\tilde{m})~,
\ee
where $A$ is $\alpha_1^2$ or $\alpha_2^2$ respectively 
depending on whether $p$ is an odd or an even integer, and $\alpha, \alpha_1, \alpha_2$ are
given by Eqs. (\ref{eq14}), (\ref{eq32}) and (\ref{eq37}) respectively;

\bea\label{f5}
\!\!\!\!\!\!\!\!\!\!\!\!\!\!\!\!\!\!\!\!&&\sum_{j=1}^{p} \dn^2 (x_j,m) \dn^2(x_{j+r},m) 
=-(2/\alpha^2) \cs^2(a,m) 
\dn^2 (x,\tilde{m}) \nonumber \\
&&~~~~~~+4A_d\cs^2(a,m)
+p\big [\cs^2(a,m)+\ds^2(a,m)-2\cs(a,m)\ds(a,m)\ns(a,m)\Z(a,m) \big ]~,
\eea
where $A_d$ is given by Eq. (\ref{eq88});

\be\label{f6}
\sum_{j=1}^{p} \dn^3 (x_j,m) [\dn(x_{j+r},m)-\dn(x_{j-r},m)] 
=-(2m/A)\cs(a,m) 
\cn (x,\tilde{m}) \sn (x,\tilde{m}) \dn (x,\tilde{m})~,
\ee
where $A$ is $\alpha\alpha_1^2$ for $p$ odd and $m\alpha \alpha_2^2$ for
$p$ even.

We now present four examples of novel Landen formulas which are only valid for odd $p$:
\be\label{f7}
m\sum_{j=1}^{p} \sn^2 (\tilde{x_j},m) 
[\sn(\tilde{x}_{j+r},m)+\sn(\tilde{x}_{j-r},m)] 
=-(2/\alpha_1)[\ds(b,m)\cs(b,m)-\ns^2(b,m)] 
\sn (x,\tilde{m})~;
\ee
\bea\label{f8}
&&\!\!\!\!\!\!\!\!\!\!\!\!\!\!\!\!\!\!\!\!m^2\sum_{j=1}^{p} \sn^3 (\tilde{x_j},m) 
[\sn^2(\tilde{x}_{j+r},m)-\sn^2(\tilde{x}_{j-r},m)] \nonumber \\ 
&&~~=(2/\alpha \alpha_1) \ns(b,m)[2\ds(b,m)\cs(b,m)+\ns^2(b,m)] 
\cn (x,\tilde{m}) \dn (x,\tilde{m})~;
\eea
\be\label{f9}
m\sum_{j=1}^{p} \cn^2 (\tilde{x_j},m) 
[\cn(\tilde{x}_{j+r},m)+\cn(\tilde{x}_{j-r},m)] 
=(2/\alpha_1)[\ns(b,m)\cs(b,m)-\ds^2(b,m)] 
\cn (x,\tilde{m})~;
\ee
\bea\label{f10}
&&\!\!\!\!\!\!\!\!\!\!\!\!\!\!\!\!\!\!\!\!m\sum_{j=1}^{p} \cn(\tilde{x_j},m) \sn(\tilde{x_j},m) \dn(\tilde{x_j},m) 
[\sn(\tilde{x}_{j+r},m) \dn(\tilde{x}_{j+r},m)
-\sn(\tilde{x}_{j-r},m) \dn(\tilde{x}_{j-r},m)] \nonumber \\
&&~~=(2/\alpha \alpha_1) \ds(b,m)[\ns(b,m)\cs(b,m)+\ns^2(b,m)+\cs^2(b,m)] 
\sn (x,\tilde{m})
\dn (x,\tilde{m})~,
\eea
where $x_j,\tilde{x_j}$ are given by Eq. (\ref{eq18}) and $b=4rK(m)/p$.

On the other hand, here are five examples of novel Landen formulas which are only valid 
for even $p$ and odd $r < p$:
\be\label{f11}
m \sum_{j=1}^{p} (-1)^{j-1} \sn(x_j,m) [\cn(x_{j+r},m)-\cn(x_{j-r},m)] = 
(2/\alpha_2)[\ns(a,m)+\ds(a,m)] 
\cn \big ({x}/{\alpha},\tilde{m} \big )~;
\ee
\bea\label{f12}
&& \!\!\!\!\!\!\!\!\!\!\!\!\!\!\!\!\!\!\!\! m\sum_{j=1}^{p} (-1)^{j-1} \sn(x_j,m) \dn(x_j,m) 
[\cn(x_{j+r},m) \dn(x_{j+r},m)+\cn(x_{j-r},m) \dn(x_{j-r},m)] \nonumber \\ 
&&~~=-(2/\alpha \alpha_2) \cs(a,m) [\ds(a,m)-\ns(a,m)] 
\sn \big ({x}/{\alpha},\tilde{m} \big )
\dn \big ({x}/{\alpha},\tilde{m} \big )~;
\eea
\be\label{f13}
\sum_{j=1}^{p} (-1)^{j-1} \dn(x_j,m) \dn(x_{j+r},m) = 
-(2/\alpha_2)\cs(a,m) 
\sn({x}/{\alpha},\tilde{m})~;
\ee
\bea\label{f14}
&&\!\!\!\!\!\!\!\!\!\!\!\!\!\!\!\!\!\!\!\!\sum_{j=1}^{p} (-1)^{j-1} \dn(x_j,m) \dn(x_{j+r},m) 
\dn(x_{j+2r},m) \dn(x_{j+3r},m) \nonumber \\ 
&&~~~=(2/\alpha_2) \big [\cs(a,m)\cs(2a,m)\cs(3a,m)+\cs^2(a,m)\cs(2a,m) \big ] 
\sn \big ({x}/{\alpha},\tilde{m} \big )~;
\eea
\be\label{f15}
 \sum_{j=1}^{p} (-1)^{j-1} \dn^3(x_j,m) [\dn(x_{j+r},m)+\dn(x_{j-r},m)] = 
(2/\alpha \alpha_2) \ns(a) \ds(a) 
\cn \big ({x}/{\alpha},\tilde{m} \big )
\dn \big ({x}/{\alpha},\tilde{m} \big )~.
\ee

\section{Conclusion}

The results of this paper clarify the relationship between the well known 
periodic solutions  
of various nonlinear differential equations and those obtained recently by us using the idea 
of judicious linear  
superposition \cite{ks1,cks}. In fact, we obtain
a deep connection between the   
highly nonlinear Landen   
transformation formulas involving changes of the modulus 
parameter and certain linear superpositions of an arbitrary 
number of Jacobi elliptic functions. Further, we have found that using 
the identities
with and without alternating signs along with Landen transformations, one can obtain
novel Landen formulas for suitable combinations of Jacobi elliptic functions.
In this context, it is worth recalling that recently we have also obtained 
cyclic identities with arbitrary weight factors and a large number of local identities \cite{kls1}.
It would be very interesting if these could  also be
profitably combined with the Landen transformations.

{\bf Acknowledgment:} 
We are grateful to the U.S. Department of Energy for providing partial support of this research under grant DOE FG02-84ER40173.

\newpage 
\bigskip 
 
\noindent {Table 1: A table showing the transformed modulus parameter ${\tilde m}$ 
appearing in the generalized Landen transformation formulas as a function of the 
modulus parameter $m$ and the number of terms $p$ in the formula. The values of ${\tilde m}$ have been computed using Eq. (\ref{eq15}). As discussed in the text, the same values of ${\tilde m}$ could equally well have been obtained from the equivalent expressions Eqs. (\ref{eq35}) or (\ref{eq53})
for odd integers $p$,
and  Eqs. (\ref{eq39}) or (\ref{eq58})
for even integers $p$. Note that the table also gives the change of modulus parameter for Gauss transformations if one interchanges the roles of $m$ and $\tilde{m}$.
\sss 
\sss 
 
\oddsidemargin      -0.3in 
\bigskip 
\begin{tabular}{ccccccc}  
\hline 
 $m$ & ${\tilde m}(p=2)$ & ${\tilde m}(p=3)$  & ${\tilde m}(p=4)$  & ${\tilde m}(p=5)$ & ${\tilde m}(p=6)$ & ${\tilde m}(p=7)$ \\ 
\hline 
 0 & 0  & 0 & 0 & 0 & 0 & 0\\ 
 0.25& .5155 x$~10^{-2}$ & .9288 x$~10^{-4}$ & .1669 x$~10^{-5}$ & .3000 x$~10^{-7}$& .5392 x$~10^{-9}$& .9693 x$~10^{-11}$\\  
 0.5& .2944 x$~10^{-1}$ & .1290 x$~10^{-2}$ & .5580 x$~10^{-4}$ & .2411 x$~10^{-5}$& .1042 x$~10^{-6}$& .4503 x$~10^{-8}$\\   
0.75& .1111 & .1005 x$~10^{-1}$ & .8666 x$~10^{-3}$ & .7438 x$~10^{-4}$& .6381 x$~10^{-5}$& .5475 x$~10^{-6}$\\   
0.9& .2699 & .4311 x$~10^{-1}$ & .6158 x$~10^{-2}$ & .8655 x$~10^{-3}$& .1213 x$~10^{-3}$& .1701 x$~10^{-4}$\\   
0.99& .6694& .2506 & .7283 x$~10^{-1}$ & .1963 x$~10^{-1}$& .5185 x$~10^{-2}$& .1362 x$~10^{-2}$\\   
0.999& .8811& .5292 & .2374 & .9312 x$~10^{-1}$& .3464 x$~10^{-1}$& .1264 x$~10^{-1}$\\   
0.9999& .9608& .7446 & .4481 & .2293& .1080 & .4891 x$~10^{-1}$\\   
0.99999 & .9874 & .8721 & .6374 & .3973 & .2239 & .1193\\   
1& 1 & 1& 1 & 1& 1& 1\\  
\hline 
\end{tabular} 
\bigskip

\end{document}